\documentclass[twocolumn,showpacs,preprintnumbers,amsmath,amssymb]{revtex4}

\usepackage{graphicx}
\usepackage{dcolumn}
\usepackage{bm}

\begin{document}
\title{Thermoelectric response near a quantum critical point: the case of CeCoIn$_{5}$ }
\author{K. Izawa$^{1,2,3}$, K. Behnia$^{4}$, Y. Matsuda$^{3,5}$, H. Shishido$^{5,6}$, R.Settai$^{6}$, Y. Onuki$^{6}$ and J. Flouquet$^{2}$}

\affiliation{$^{1}$Department of Physics, Tokyo Institute of Technology, Meguro, Tokyo, 152-8551 Japan}
\affiliation{$^{2}$DRFMC/SPSMS,  Commissariat \`a l'Energie Atomique,
F-38042 Grenoble, France}
\affiliation{$^{3}$Institute for Solid State Physics, University of Tokyo, Kashiwa, Chiba 277-8581, Japan}
\affiliation{$^{4}$Laboratoire Photons Et Mati\`ere(CNRS), ESPCI, 
75231 Paris, France}
\affiliation{$^{5}$Department of Physics, Kyoto University, Kyoto 606-8502, Japan}
\affiliation{$^{6}$Department of Physics, Osaka University, Toyonaka, Osaka
560-0043, Japan}


\begin{abstract}
We present a study of thermoelectric coefficients in CeCoIn$_{5}$
down to 0.1~K and up to 16~T in order to probe the thermoelectric
signatures of quantum criticality. In the vicinity of the
field-induced quantum critical point, the Nernst coefficient $\nu$
exhibits a dramatic enhancement without saturation down to lowest
measured temperature. The dimensionless ratio of Seebeck coefficient
to electronic specific heat shows a minimum at a temperature close
to threshold of the quasiparticle formation. Close to $T_{c}(H)$, in
the vortex-liquid state, the Nernst coefficient behaves anomalously
in puzzling contrast with other superconductors and standard vortex
dynamics.

\end{abstract}


\maketitle

CeCoIn$_{5}$ is an unconventional superconductor with an intriguing
normal state\cite{petrovic}. Its behavior is peculiar near the upper
critical field, where the  energy scale governing various electronic
properties is vanishingly small and increases with increasing
magnetic field\cite{paglione,bianchi}, a behavior expected in
presence of a Quantum Critical Point(QCP)\cite{stewart}. The
proximity of this QCP to the upper critical field in CeCoIn$_{5}$ is
puzzling\cite{bauer,ronning,paglione2}. 
The possible existence of a FFLO state\cite{fflo} and/or an elusive magnetic order are a subject of recent intense research. 
On the other hand, even in the absence of magnetic field, the normal state
presents strong deviation from the standard Fermi-liquid
behavior\cite{petrovic,nakajima}. The application of pressure leads
to the destruction of superconductivity and the restoration of the
Fermi liquid\cite{sidorov,ronning,nakajima2}. The link between the
field-induced and the pressure-induced routes to the Fermi liquid is
yet to be clarified.

During the last three years, the anomalous properties of
CeCoIn$_{5}$ near the field-induced QCP  have been 
reported 
thanks to measurements of specific heat\cite{bianchi},
electric resistivity\cite{paglione}, thermal
transport\cite{paglione2} and Hall effect\cite{singh}. 
In this paper, new insight on the quantum criticality is given via thermoelectric response down to 0.1K. 
As far as we know, this is the first experimental
investigation of the thermoelectric tensor in the vicinity of a QCP,
a subject of several theoretical studies\cite{paul,miyake,podolsky}.
Single crystals were grown by self-flux method. Thermoelectric coefficients were measured with one heater and two RuO$_{2}$ thermometers in magnetic field along $c$-axis. The heat current was applied along the basal plane. 
Previous studies of thermoelectricity in
CeCoIn$_{5}$ detected a large Nernst
coefficient and a field-dependent Seebeck coefficient in the
non-Fermi liquid regime above $T_{c}$\cite{bel} and an additional
field scale at 23~T\cite{sheikin}. 
Here, we find that the most spectacular thermoelectric signature of
quantum criticality is a drastic enhancement of the Nernst
coefficient, $\nu$.  The vanishingly small Fermi energy, which was
previously detected by a nearly diverging enhancement of the
$A$ coefficient of resistivity ($\rho=\rho_{0}+AT^{2}$)\cite{paglione} and
the Sommerfeld coefficient of specific heat ($\gamma=C_{el}/T$)\cite{bianchi},
leads also to an apparently diverging $\nu/T$. 
These results show two distinct anomalies close to $H_{c2}(0)$ and $T_{c}(0)$ which are different in the origin. This conclusion cannot be derived from other probes mentioned above. 
We also find a milder
enhancement of the Seebeck coefficient near the QCP. Moreover, the
ratio of thermopower to electronic specific heat, expressed in
appropriate units\cite{behnia}, remains close to unity even in the
vicinity of the QCP. The temperature dependence of this ratio
presents a minimum at a temperature roughly marking the formation of
well-defined quasi-particles\cite{paglione2}.

Figure 1 presents the  data obtained by measuring the Nernst and the
Seebeck coefficients at various magnetic fields. Since the
thermoelectric response of Fermions is expected to be $T$-linear well
below their Fermi temperature, what is plotted in the figure is the
temperature dependence of the two coefficients divided by
temperature. As seen in fig.~1(a), the Seebeck coefficient, $S$
vanishes in the superconducting state. In the normal state, $S/T$
increases with decreasing temperature for all fields.  For fields
exceeding 5.4~T, the normal state extends down to zero temperature
and a finite $S/T$ in the zero-temperature limit can be extracted.
For a field of 16~T (which is well above the quantum critical region)
$S/T$ saturates to a value of about 13~$\mu$VK$^{-2}$. For fields
between 5.4~T and 16~T, $S/T$ presents a non-monotonous temperature
dependence. An upturn below 0.15~K is visible for $\mu_{0}H \simeq$ 5.5~T
(i.e. in the vicinity of the QCP) curves. Note that this upturn
leads to a moderate enhancement of $S/T$. The overall change in the
magnitude of $S/T$ is about 70~\%. On the other hand, the
temperature dependence of the Nernst coefficient divided by
temperature $\nu/T$ reveals a more dramatic signature of Quantum
criticality. As seen in figure 1(b), for $\mu_{0}H=$ 5.5~T and $\mu_{0}H=$ 6~T,
below 1~K, $\nu/T$ is steadily increasing with decreasing
temperature. No such enhancement occurs for $\mu_{0}H=$16~T, far above QCP.
At the lowest measured temperature ($\sim$ 0.1~K), $\nu/T$ is
five-fold enhanced near the QCP ($\sim$6~T) compared to its 16~T
value. Since the thermal Hall conductivity $\kappa_{xy}$ in
CeCoIn$_{5}$ becomes large at low temperatures due to enhancement of
the mean-free-path of the electrons \cite{kasahara}, the transverse
thermal gradient $\nabla_{y} T$ could generate a finite
transverse electric field $E_{y}$. Therefore, the (measured)
adiabatic and the (theoretical) isothermal Nernst coefficients are
not identical in CeCoIn$_{5}$. However, using the value of
$|\nabla_{y} T|/|\nabla_{x} T| \sim$ 0.1 at 5.2~T reported in Ref.~\cite{kasahara}, the difference between these two is
estimated to be about 10~\%, indicating that the observed
enhancement is not due to a finite $\nabla_{y} T$. We will argue
below that this enhancement reflects a concomitant decrease in the
magnitude of the normalized Fermi energy as previously documented by
specific heat and resistivity measurements.
\begin{figure}
\resizebox{!}{0.23\textwidth}{\includegraphics{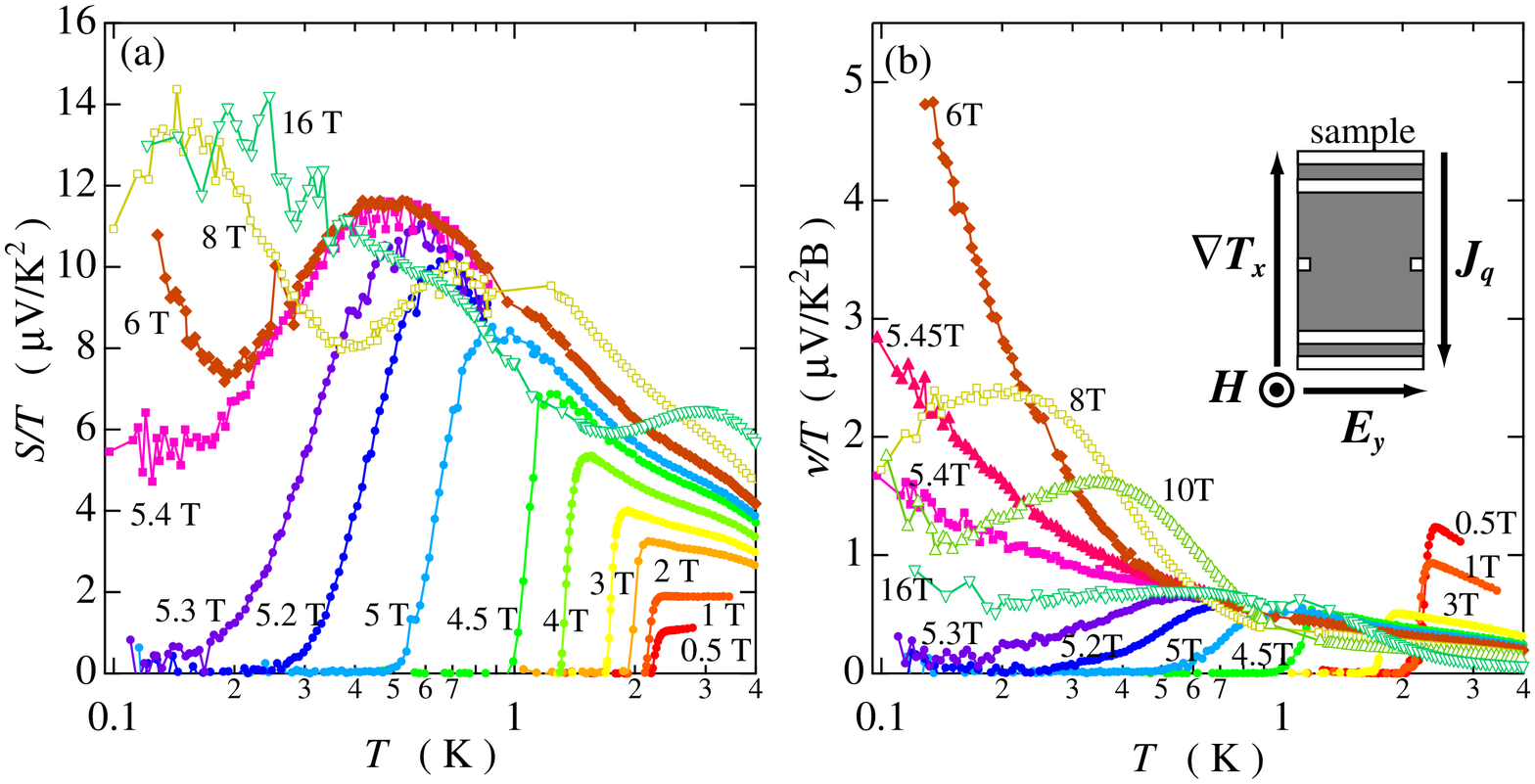}}
\caption{\label{fig1}(a) The Seebeck coefficient divided by
temperature  as a function of temperature for different magnetic
fields in a semi-log plot. Note the upturn near the QCP.(b)
Temperature dependence of the Nernst coefficient divided by
temperature $\nu/T$. Close to the QCP, this quantity never
saturates. The inset defines the convention used for the sign of the
Nernst coefficient(see text).}
\end{figure}

The thermoelectric response of CeCoIn$_{5}$ in the vicinity of QCP
can be better understood by complementing our data with the
information extracted by other experimental
probes\cite{bianchi,paglione}, which originally detected a quantum
critical behavior near $H_{c2}$. In particular, an interesting issue
to address is the fate of the correlation observed between
thermopower and specific heat of many Fermi liquids in the
zero-temperature limit
 In a wide range of systems, the
dimensionless ratio linking these two is of the order of unity
($q=\frac{S N_{A} e}{T\gamma}\simeq\pm 1$, with $\gamma=C_{el}/T$,
 $N_{A}$ the Avogadro number and $e$ the
charge of electron)\cite{behnia}. What happens to such a correlation at a quantum
critical point? 
Combining the specific heat data reported by Bianchi \emph{et
al.}\cite{bianchi} with our thermopower results allows us to address
these questions. Fig.~2(a) presents $q$ computed in this way as a
function of temperature. 
The first feature to remark is that $q$
remains of the order of unity even in the quantum critical region.
Note that, theoretically, this correlation arises because $S/T$ and
$\gamma$ are both inversely proportional to the normalized Fermi
energy and thus $q$ is expected to be of the order of (and $not$
rigorously equal to) unity\cite{miyake}. According to our result
($q\simeq 0.9$ at 6~T and 0.1~K), this correlation holds even when
the normalized Fermi Energy becomes vanishingly small. The second
feature of interest in figure 2(a) is the temperature dependence of
$q$, which presents a minimum. For both fields, the temperature at
which this minimum occurs is close to the one where the Lorenz
number($L=\frac{\kappa}{\sigma T}$) linking thermal, $\kappa$, and
electric,$\sigma$, conductivities present also a minimum. Paglione
and co-workers, who report this latter feature, argue that this
temperature marks the formation of well-defined
quasi-particles\cite{paglione2}. This is a temperature below which
both thermal and electric resistivities display a $T^{3/2}$
temperature dependence. Remarkably, Miyake and Kohno, who provided a
theoretical framework in a periodic Anderson model for the
correlation between thermopower and specific heat, predicted that
$q$ should deviate downward from unity in presence of an
antiferromagnetic (AF) QCP leading to hot lines on the Fermi
surface\cite{miyake}.

We now turn to the Nernst coefficient. In a simple picture, it is
proportional to the energy derivative of the relaxation time at the
Fermi energy\cite{sondheimer}. In a first approximation, it tracks a
magnitude set by the cyclotron frequency, the scattering time and
the Fermi energy\cite{behnia2}. Since it scales inversely with the
Fermi Energy, there is no surprise that it becomes large in
heavy-fermion metals\cite{bel,sheikin} and in particular in
heavy-Fermion semi-metals\cite{bel2,pourret}, where both the heavy
mass of electrons and the smallness of the Fermi wave-vector
contribute to its enhancement($\nu/T \propto
1/(k_{F}\epsilon_{F}$)). Now, since the Fermi energy (broadly
defined as the characteristic energy scale of the system) becomes
very small near a QCP, one would expect a large Nernst coefficient
in agreement with the experimental observation reported here.

\begin{figure}[t]
\resizebox{!}{0.22\textwidth}{\includegraphics{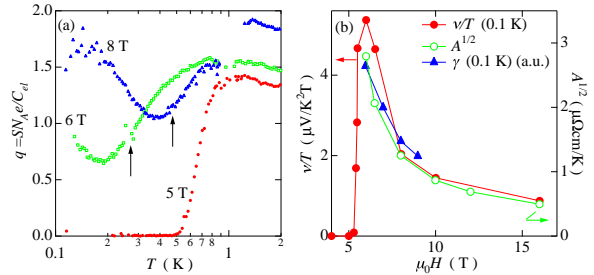}}
\caption{\label{fig2}(a) The temperature dependence of $q$, the
dimensionless ratio of thermopower to electronic specific heat at
three magnetic fields. The temperature marked by the arrow designs
the threshold of the quasi-particle formation according to the
temperature dependence of the Lorenz number as reported by Paglione
\emph{et al.}\cite{paglione2} (b) A comparison of the field
dependence of $\nu/T$, $\gamma$ (as reported in ref.~\cite{bianchi})
and $A^{1/2}$ (taken from ref.~\cite{paglione}). }
\end{figure}

With these phenomenological considerations in mind let us compare the behavior of the Nernst coefficient with specific heat and resistivity. 
Both $\gamma$ and $A$, the $T^2$ term of the
resistivity ($\rho=\rho_{0}+AT^{2}$) inversely scale with the Fermi
Energy, $\epsilon_{F}$. Therefore, both are enhanced when the Fermi
energy is small. Since these two quantities are linked by the
Kadowaki-Woods relation ($\gamma^{2}\propto A $), the enhancement is
more pronounced in $A$ than in $\gamma$. Figure 2(b) compares the
field-dependence of $A^{1/2}$, $\gamma$ and $\nu/T$. In a naive
picture, 
the enhancement of the
three quantities are comparable in magnitude. This
quantitative correlation suggests that the main reason for the
enhancement of $\nu/T$ near QCP is due to a small $\epsilon_{F}$.
%
It
is instructive to trace a contour plot of this quantity in the
temperature-field plane. This is done in Fig.~3 with a logarithmic
color scale in order to enhance the contrast. 
Note that 
contrary to the other probes, there is no need to subtract an offset from the Nernst data. 
In the case of specific heat,
one should subtract the Schottky contribution at low
temperature\cite{bianchi} and high-field, and the phonon contribution
at high temperature. In the case of resistivity the $T^2$
behavior is interrupted at low temperature and high-fields by an
upturn due to the temperature-dependent magnitude of
$\omega_{c}\tau$\cite{paglione}. As seen in Fig.~3, $\nu/T$ becomes
very large near the QCP, which constitutes the main hearth of the
figure. However, there is a second one at zero field just above
$T_c$, which was identified by previous measurements\cite{bel}.
This zero-field hot region corresponds to a purely linear resistivity and anomalously enhanced Hall coefficient\cite{nakajima} due to 
strong anisotropic scattering by AF fluctuations\cite{nakajima2},
which can also enhance the Nernst coefficient\cite{kontani}. On the
other hand, close to the QCP, the magnitude of Hall
coefficient\cite{singh} is comparable to its value at
room-temperature or in LaCoIn$_{5}$\cite{nakajima2}. Therefore,
there appears to be two distinct sources for the enhancement of the
Nernst coefficient. In the zero-field regime just above $T_c$, it is
enhanced mostly because of strong inelastic scattering associated
with AF fluctuations, but in the zero-temperature regime just above
$H_{c2}$, it becomes large because of the smallness of the Fermi
energy. The occurrence of superconductivity impedes to explore the
route linking together these two hot regions of the (B,T) plane. The
inset of the figure  compares the evolution of energy scales
detected by different experimental probes near the QCP.

\begin{figure}[t]
\resizebox{!}{0.32\textwidth}{\includegraphics{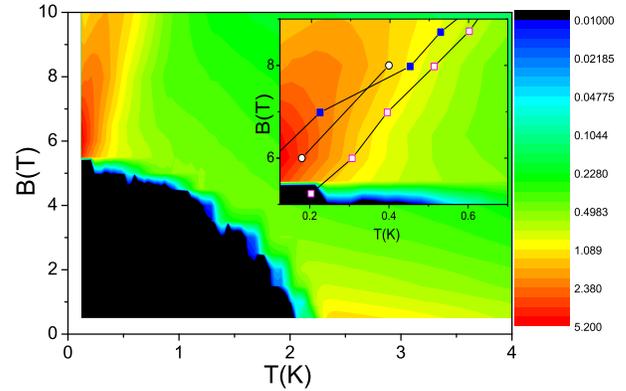}}
\caption{\label{fig3} Contour plot of $\nu/T$ in the (B,T) plane.
The color scale is logarithmic. Note the presence of two hot regions
close to $H_{c2}$ and $T_{c}$. The inset is a zoom on the region
near $H_{c2}$. The variation of three temperature scales, the onset
of $T^2$ resistivity(solid squares), the minimum in $L/L_{0}$(open
squares) the and minimum in $q$(open circles) with magnetic field is
also shown.}
\end{figure}

We now turn to the puzzling behavior of the Nernst coefficient in
the vicinity of the superconducting transition. Deep into the
superconducting state, there is no measurable Nernst signal, as
illustrated by the existence of the black area in Fig.~3. On the
other hand, close to $H_{c2}$(T) (or alternatively, near $T_{c}(H)
$), vortices can move and an additional contribution to the Nernst
signal is expected. In the entire range of our study, the Nernst
coefficient keeps the same sign which is presented in the inset of
Fig.~1. Such a Nernst coefficient is negative according to a
textbook convention on the sign of the thermoelectric
coefficients\cite{nolas}. However, the literature on the vortex
Nernst effect\cite{wang} usually takes for positive the Nernst
signal generated by vortices moving from hot to cold, which leads to
an opposite convention. The sign of the Nernst effect in
CeCoIn$_{5}$ is negative according to the textbook
convention\cite{nolas}, but positive according to the vortex
one\cite{wang,huebener}. Indeed, contrary to quasi-particles, the
Nernst signal produced by vortices should have a fixed sign. A
thermal gradient $\nabla_{x} T$ generates a force on a vortex
because its core has an excess of entropy. The direction of this
force is thus thermodynamically determined; vortices move along the
thermal gradient from hot to cold region. The orientation of
electric field is also unambiguously set by the direction of the
vortex movement and the vortex Nernst signal is not expected to have
an arbitrary sign. In order to separate the vortex and the
quasi-particle contributions to the Nernst signal, we put under
careful scrutiny the effect of superconducting transition on three
coefficients : $\rho(T)$, $S(T)$ and $N(T)$ . As illustrated in
fig.~4(a) and 4(b), with the onset of superconductivity, the Nernst
signal, $N$, collapses faster than both resistivity and the Seebeck
coefficient. This robust feature was observed for \emph{all}
magnetic fields. On the other hand, the collapse in $\rho(T)$ and
$S(T)$ closely track each other. This latter feature, which was also
observed in cuprates\cite{huebener}, suggests that the Seebeck
response is essentially generated by quasi-particles. Therefore, the
most natural assumption regarding their contribution to the Nernst
signal in the vortex liquid regime is  that $N_{qp}(T)$ also follows
$\rho(T)$ and $S(T)$ and the vortex contribution to the Nernst
signal can be obtained by subtracting the normalized Seebeck
coefficient off the normalized Nernst one. Fig.~4(c) and 4(d) show
that this procedure clearly resolves a signal of opposite sign.
Thus, the most straightforward interpretation of the faster collapse
of $N(T)$ implies an additional source of Nernst signal in the
vortex liquid regime with a sign \emph{opposite} to the predominant
one and also to the one expected for vortices moving along the heat
flow.

This result appears incompatible with the standard picture of vortex
dynamics driven by a thermal gradient. However, one shall not forget
that additional forces on vortices besides thermal force may be
present. CeCoIn$_{5}$ is distinguished from other superconductors by
the possible occurrence of an anti-ferromagnetic state in the normal
core of its vortices. This feature could decrease the entropy excess
of the vortices and reduce the intensity of the thermal force, which
can therefore be vanquished by another source of vortex movement. As
first noted by Ginzburg\cite{ginzburg}, in a superconductor subject
to a thermal gradient, a quasi-particle current (which carry heat)
and a supercurrent (which does not) counterflow in order to keep the
charge current zero\cite{huebener}. In ordinary conditions, this
counterflow generates a transverse Magnus force on
vortices\cite{ri}. Its role in the context of superclean
CeCoIn$_{5}$\cite{kasahara} needs an adequate theoretical
treatment.

\begin{figure}[t]
\resizebox{!}{0.32\textwidth}{\includegraphics{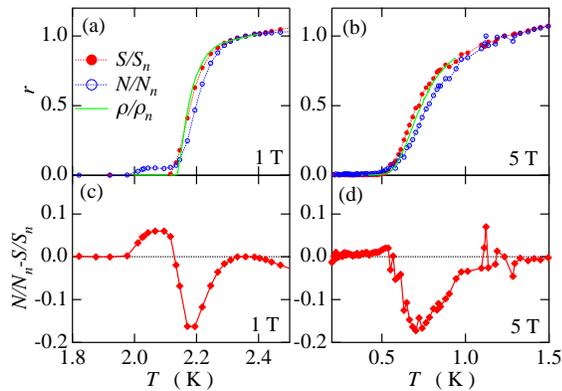}}
\caption{\label{fig4}(a)(b) Normalized magnitudes $r$ of $r=N/N_{n}$(open
circles), $S/S_{n}$(solid circles) and $\rho/\rho_{n}$(solid line) in the vicinity
of superconducting transition for 1~T and 5~T. For all fields, the
onset of superconductivity leads to a faster collapse of the Nernst
signal. Note also the small shoulder at 1~T. (c)(d) The additional
contribution to transverse thermoelectricity in the vortex liquid
regime obtained by subtracting the normalized Seebeck coefficient
off the normalized Nernst signal. }
\end{figure}

Another remarkable feature of Fig.~4 is the presence of a small
shoulder in the temperature dependence of the Nernst effect at the
end of the transition. The shoulder is present in an extended range
of magnetic fields and only disappears in the proximity of $H_{c2}$.
There seems to be a narrow temperature window, where a thermal
gradient can create a transverse electric field, but a current does 
not produce any electric field. The simplest explanation for such a
discrepancy would imply a threshold force to depin vortices,
$f_{dp}$  attained by the applied temperature gradient, but not by
the applied current. 
However, this feature was found to be robust and no change was
detected by modifying the magnitude of the applied thermal gradient.
Clearly, the sign and the fine structure of the Nernst effect in the
vortex liquid regime of CeCoIn$_{5}$ need further investigation.

We thank J-P. Brison, H. Kontani, N. Kopnin, K. Maki and K. Miyake for helpful discussions and  specially N. P. Ong for his illuminating input on the sign of the Nernst effect. K.I. acknowledges a European Union Marie Curie fellowship. This work was supported by the Agence Nationale de la Recherche through the ICENET project.


\begin{thebibliography}{99}
\bibitem{petrovic} C. Petrovic \emph{et al.}, J. Phys. Condens. Matter \textbf{13}, L337 (2001).
\bibitem{paglione} J. Paglione \emph{et al.},Phys. Rev. Lett. \textbf{91}, 246405 (2003).
\bibitem{bianchi} A. Bianchi \emph{et al.}, Phys. Rev. Lett. \textbf{91}, 257001 (2003).
\bibitem{stewart} G. R. Stewart, Rev. Mod. Phys. \textbf{73}, 797 (2001).
\bibitem{bauer} E. D. Bauer \emph{et al.}, Phys. Rev. Lett. \textbf{94}, 047001 (2005).
\bibitem{ronning} F. Ronning \emph{et al.}, Phys. Rev. B \textbf{73}, 064519 (2006).
\bibitem{paglione2} J. Paglione \emph{et al.},Phys. Rev. Lett. \textbf{97}, 106606 (2006).
\bibitem{fflo} Y. Matsuda and H. Shimahara, J. Phys. Soc. Jpn \textbf{76}, 051005 (2007). 
\bibitem{nakajima} Y. Nakajima \emph{et al.}, J. Phys. Soc. Jpn. \textbf{73}, 5 (2004).
\bibitem{sidorov} V. A. Sidorov \emph{et al.}, Phys. Rev. Lett. \textbf{89}, 157004 (2002).
\bibitem{nakajima2} Y. Nakajima \emph{et al.}, J. Phys. Soc. Jpn. \textbf{76}, 024703 (2007).
\bibitem{kasahara} Y.Kasahara {\it et al.}, Phys. Rev. B {\bf 72}, 214515 (2005).
\bibitem{singh} S. Singh \emph{et al.}, Phys. Rev. Lett. \textbf{98}, 057001 (2007).
\bibitem{bel} R. Bel \emph{et al.}, Phys. Rev. Lett. \textbf{92}, 217002 (2004).
\bibitem{sheikin} I. Sheikin \emph{et al.}, Phys. Rev. Lett. \textbf{96}, 077207 (2006).
\bibitem{paul} I. Paul and G. Kotliar, Phys. Rev. B \textbf{64}, 184414 (2001).
\bibitem{miyake}  K. Miyake and Kohno,  J. Phys. Soc. Jpn.  \textbf{74}, 254 (2005).
\bibitem{podolsky}D. Podolsky \emph{et al.}, Phys. Rev. B \textbf{75}, 014520 (2007).
\bibitem{behnia} K. Behnia \emph{et al.}, J. Phys.: Condens. Matter \textbf{16}, 5187 (2004).
\bibitem{sondheimer} E. H. Sondheimer, Proc. R. Soc. London, Ser. A 193, \textbf{484}(1948).
\bibitem{behnia2} K. Behnia \emph{et al.}, Phys. Rev. Lett. \textbf{98}, 076603 (2007).
\bibitem{bel2} R. Bel \emph{et al.},  Phys. Rev. B \textbf{70}, 220501(R) (2004).
\bibitem{pourret} A. Pourret \emph {et al.} Phys. Rev. Lett. \textbf{96}, 176402 (2006).
\bibitem{kontani}H. Kontani, Phys. Rev. Lett. \textbf{89}, 237003 (2002).
\bibitem{nolas} G. S. Nolas \emph{et al.}, Thermoelectrics, Springer (2001).
\bibitem{wang}Y. Wang \emph{et al.}, Phys. Rev. B \textbf{73}, 024510 (2006).
\bibitem{huebener}R. P. Huebener, Supercond. Sci. Technol. \textbf{8}, 189
(1995).
\bibitem{ginzburg}V. L. Ginzburg, Sov. Phys. Usp. \textbf{34}, 101(1991).
\bibitem{ri} H. -C. Ri \emph{et al.}, Phys. Rev. B \textbf{47}, 12312 (1993).
\end{thebibliography}
\end{document}